# BLACK HOLES

## Affiliation

Mar Mezcua

Institute of Space Sciences (ICE, CSIC), Campus UAB, Carrer de Magrans, 08193 Barcelona, Spain
marmezcua.astro@gmail.com

## Acronyms

$M_{BH}$ (black hole mass), SMBH (supermassive black hole), IMBH (intermediate-mass black hole)

## Definition

Black holes are defined as a region in spacetime where gravity is so strong that particles and electromagnetic radiation cannot escape. At the center of a black hole density and gravity become infinite, in what is called the singularity. The boundary region within nothing can escape is called the event horizon. The radius of this event horizon, which determines the size of a black hole, is called Schwarzschild radius and is directly proportional to the black hole mass ($M_{BH}$): $R_{Sch}=2GM_{BH}/c^2$ where G is the gravitational constant and c the speed of light (see Fig. 1). Mass, spin and charge are the three properties that characterize black holes. A non-rotating black hole with no charge is called Schwarzschild black hole while a rotating one is called Kerr black hole (see History section). At a distance of 1.5 times the Schwarzschild radius of a non-rotating black hole there is a spherical boundary where photons moving on tangential trajectories to that sphere are trapped around the black hole in circular orbits. For a Kerr black hole this photon sphere has a radius that depends

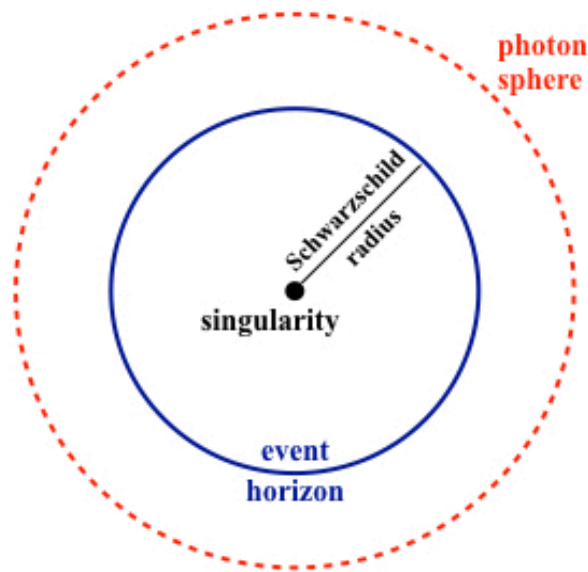

Fig. 1. Anatomy of a Schwarzschild black hole. The event horizon is spherical and its radius is called Schwarzschild radius ($R_{Sch}$). At $1.5R_{Sch}$ there is the photon sphere inside which photons move around the black hole in circular orbits.

on the spin of the black hole and the orbit of the photon.

# History

Black holes were proposed by Karl Schwarzschild in 1916 to solve the Einstein equations of general relativity. Schwarzschild's metric described a non-rotation black hole with no charge (Schwarzschild black hole). The solution for a rotating black hole was found in 1963 by Roy Kerr (Kerr black hole), and that of a rotating black hole electrically charged by Ezra Newman in 1965. Several observational techniques have since then provided evidence for the existence of black holes (e.g. the detection of signatures of black hole accretion, measurements of stellar and gas kinematics within the radius of influence of a black hole; see Basic Methodology section). However, it wasn't until 2016 that the detection of gravitational waves provided direct evidence that black holes exist and merge (Abbott et al. 2016).

# Overview

According to their mass, black holes are classified into three types:

- **Stellar-mass black holes.** They have a mass in the range $3\ M_\odot < M_{BH} \leq 100\ M_\odot$ and form when a massive star ($> 15\ M_\bullet$) burns out its fuel supply and its internal pressure is no longer able to support gravitational collapse. Most stellar-mass black holes have been detected thanks to being part of an X-ray binary system in which the black hole accretes matter from a stellar companion via an accretion disk. The detection of X-ray and optical emission from X-ray binaries in the 1970s and 1980s (e.g. Cygnus X-1, LMC-X3; see review by e.g. Remillard & McClintock 2006) provided the first observational evidence for the existence of stellar-mass black holes with a mass above $3\ M_\odot$ (i.e. too large for a white dwarf or a neutron star). Since then, more than 20 stellar-mass black holes have been confirmed in X-ray binaries (e.g. Casares & Jonker 2014). The advent of gravitational wave interferometers such as LIGO and Virgo has provided the detection of tens of additional stellar-mass black holes (see Fig. 2).

- **Supermassive black holes (SMBHs).** They have a mass $M_{BH} \geq 10^6\ M_\odot$ and reside typically at the center of massive galaxies (stellar mass $M_* \geq 10^{10}\ M_\bullet$). SMBHs of up to $\sim 10^9\ M_\odot$ in the local Universe are thought to result from the growth through cosmic time of stellar-mass black holes of a few tens of solar masses via accretion and mergers with other black holes. The detection of SMBHs of up to $\sim 10^{10}\ M_\odot$ when the Universe was less than 1 Gyr old (redshift z~7; e.g. Wang et al. 2021) suggests they have to grow instead from higher-mass seed black holes of $100\ M_\odot < M_{BH} < 10^6\ M_\odot$ at z > 10 in order to reach $10^{10}\ M_\odot$ in such a short time (see Fig. 3). The best observational evidence for the existence of SMBHs comes from the measurement of stellar motions around the Galactic center, which revealed the presence of a SMBH of $4 \times 10^6\ M_\odot$ (Ghez et al. 2008; Gillessen et al. 2009). More recently, the Event Horizon Telescope collaboration provided the first event-horizon-scale image of (the shadow of) a SMBH in the massive elliptical galaxy M87 (The EHT Collaboration et al. 2019). A SMBH that is actively accreting matter from its surrounding is called an Active Galactic Nucleus (AGN). An AGN can radiate over a trillion times the energy of the Sun, outshining the galaxy in which it resides.

- **Intermediate-mass black holes (IMBHs).** They have a mass $100\ M_\bullet_\odot < M_{BH} < 10^6\ M_\odot$ and are thought to be the seeds from which SMBHs in the early Universe grew (see reviews by Mezcua 2017; Greene et al. 2020). Such seed IMBHs could form at z > 10 from the death of Population III stars of more than 260 $M_\odot$, from direct collapse of pristine gas, or from runaway mergers in dense stellar clusters, among other possibilities (Rees 1978). They could then rapidly grow via accretion and mergers to become the SMBHs observed by z~7 (see Fig. 3; Mezcua 2017 and references therein). Those seed black holes that did not grow should be found in the local Universe as leftover IMBHs, with a mass of $\sim 10^2$-$10^3\ M_\odot$ if they originated from Population III stars, of up to $10^4\ M_\odot$ if they originated in nuclear stellar clusters, or of $\sim 10^4$-$10^6\ M_\odot$ if they formed via direct collapse of pre-galactic gas. Hundreds of IMBH candidates have-been found in the local Universe (e.g. Reines et al. 2013; Chilingarian et al. 2018; Mezcua & Dominguez Sanchez 2020) and out to z~3.4 (Mezcua et al. 2016, 2018b, 2019) as low-mass ($M_{BH} < 10^6$ M☉) AGN in dwarf galaxies. Because of their low mass (stellar mass $M_* \leq 3 \times 10^9$ M☉) and quiet merger history dwarf galaxies are thought to resemble the first galaxies of the early Universe. Tens of additional candidates have been also found in globular clusters (e.g. Luetzgendorf et al. 2013; see Table 1 in Mezcua 2017 and references therein) and in the outskirts of massive galaxies that have recently undergone a merger with a dwarf galaxy (e.g. Mezcua et al. 2015, 2018c; Barrows et al. 2019). The best observational evidence for the existence of IMBHs was provided by the LIGO/Virgo detection of a local IMBH of 142 M☉ (Fig. 2; Abbot et al. 2020).

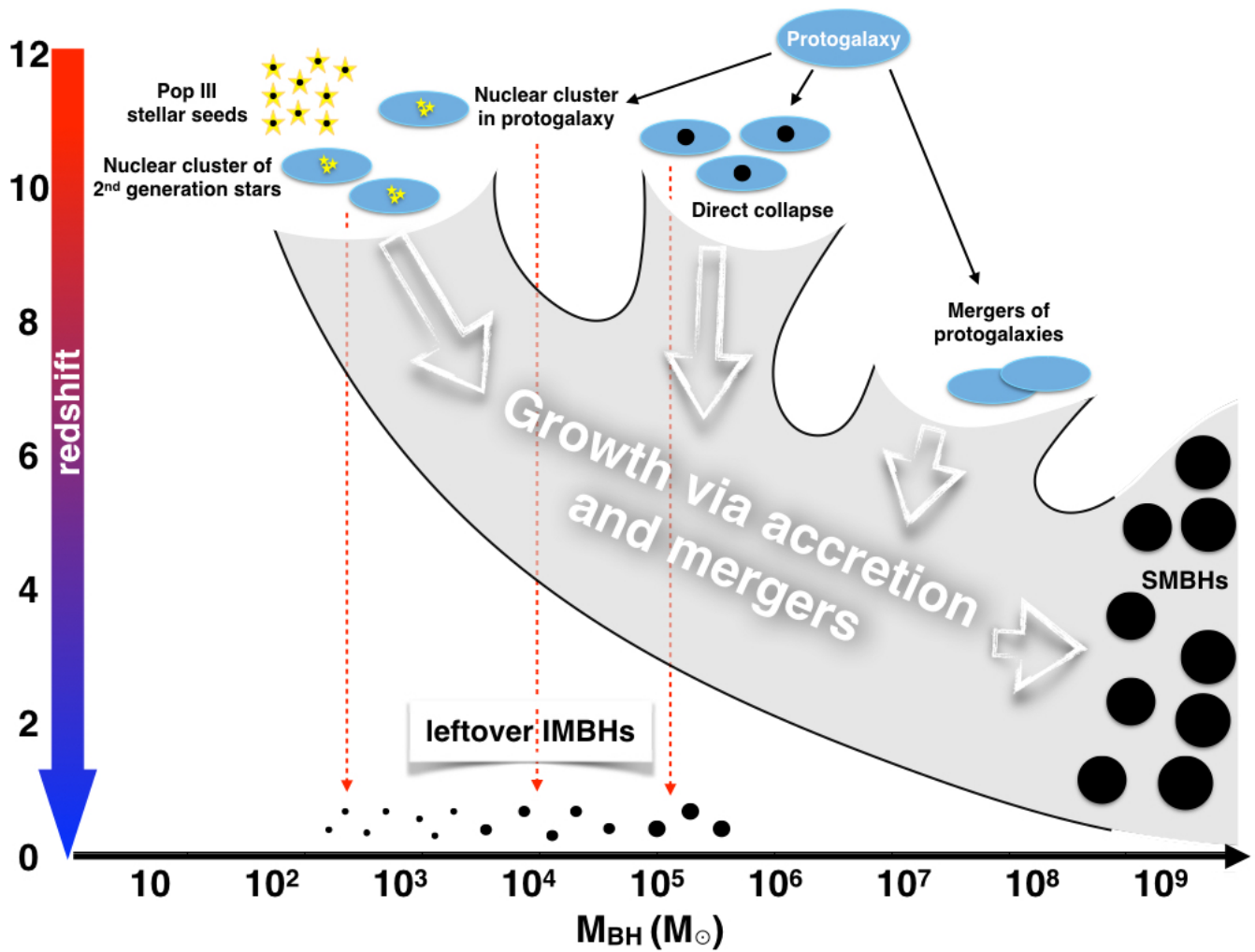

Fig. 3. Formation scenarios for IMBHs. Seed black holes in the early Universe could form from Population III stars, from mergers in dense stellar clusters formed out either from the second generation of stars or from inflows in protogalaxies, or from direct collapse of dense gas in protogalaxies, and grow via accretion and merging to $10^9$ M☉ by redshift z~7. SMBHs could also directly form by mergers of protogalaxies at z~6. Those seed black holes that did not grow into SMBHs can be found in the local Universe as leftover IMBHs. Credit: Mezcua (2017), International Journal of Modern Physics D, 26, 11, reproduced with permission.

A possible forth type of black holes are primordial black holes. These could have formed in the presence of high-density fluctuations in the inflationary Universe or of abrupt changes in the plasma pressure, and could have masses ranging from sub-solar values to $10^8$ M☉ (e.g. Rubin et al. 2001; Carr et al. 2020; Volonteri et al. 2021). Primordial black holes are currently being considered as a form of dark matter and for interpretation of gravitational wave events and, if sufficiently large, they can also be regarded as pregalactic seeds of SMBHs (e.g. Rubin et al. 2001; García-Bellido 2019; Carr et al. 2020).

## Basic Methodology

There are several ways to weight and detect black holes.

**Dynamical mass measurements.**

In the case of stellar-mass black holes, most black hole masses come from dynamical measurements based on the detection of the radial velocity curve of the donor star in X-ray binaries (see review by e.g. Casares & Jonker 2014). Other (indirect) methods, used to estimate the mass of a few IMBH candidates, include the detection of quasi-periodic oscillations (QPOs) and applying an inverse scaling relation found for stellar-mass black holes between QPO frequency and black hole mass (e.g. Pasham et al. 2014).

In globular clusters, the black hole mass of several IMBH candidates has been derived based on stellar kinematics and surface brightness profiles combined with dynamical modeling (e.g. Gebhardt et al. 2005; Luetzgendorf et al. 2013). However, the absence of any signatures of accretion in the form of X-ray or radio emission precludes confirming the presence of IMBHs in these sources (e.g. Wrobel et al. 2020).

In the case of SMBHs, the best black hole mass measurements come from the monitoring of stellar proper motions around the black hole (e.g. Ghez et al. 2008; Gillessen et al. 2009) and from dynamical modeling of stellar and gas kinematics within the black hole sphere of influence (see review by e.g. Kormendy & Ho 2013). Such dynamical mass measurements have been also possible for a few nearby IMBHs (e.g. Davis et al. 2020). Beyond the local Universe (the Local Group in the case of IMBHs), where spatial resolution prevents performing such dynamical mass measurements, the detection of IMBHs and SMBHs has to rely on the finding of signatures of accretion in the form of AGN.

**Active galactic nuclei (AGN).**

AGN can be identified across the electromagnetic spectrum. In the optical and infrared regime, the presence of AGN ionization can be distinguished from ionization by galactic star formation using emission line diagnostic diagrams. AGN can be also identified by the detection of broad emission lines coming from gas clouds moving at velocities of ~3000 km/s around the central SMBH (the so-called 'broad line region'). In the case of low-mass AGN in dwarf galaxies, these velocities can be of just hundreds of km/s (e.g. Reines et al. 2013). Such low velocities can be mimicked by some varieties of supernovae; however, unlike AGN, the broad line emission in supernovae would significantly decrease or disappear over a timespan of 5-15 years (e.g. Baldassare et al. 2016). In AGN with broad emission lines, the flux and width of the line can be used to estimate the black hole mass under the assumption that the gas in the broad line region is virialised.

In the mid-infrared regime, color-color diagrams can be also used to identify AGN (e.g. Stern et al. 2012). These diagrams are based on the colors that dust presents when heated by different mechanisms such as AGN or stars.

The detection of hard (2-10 keV band) X-ray emission from the corona of electrons sitting above the accretion disk around a black hole is a very strong tracer of black hole accretion. In massive galaxies, AGN can be identified as having an X-ray luminosity typically above $10^{42}$ erg/s. In dwarf galaxies, the AGN X-ray luminosity can be as low as $10^{38-39}$ erg/s, probing very low rates of accretion (e.g. Gallo et al. 2010). In this case, the contribution from X-ray binaries to the X-ray emission must be removed in order to confirm the presence of AGN (e.g. Mezcua et al. 2016, 2018b).

Some AGN eject a pair of collimated jets of relativistic plasma that radiate across the electromagnetic spectrum via synchrotron emission. These jets are best observed at radio wavelengths, and their detection is a common tracer of AGN activity. Radio luminosities at 1.4 GHz above $10^{24}$ W/Hz are typically used to identify AGN, as these high radio luminosities cannot be explained even by extreme star formation. In the case of dwarf star-forming galaxies, the radio contribution from star formation processes must be removed in order to infer the presence of AGN (e.g. Mezcua et al. 2019).

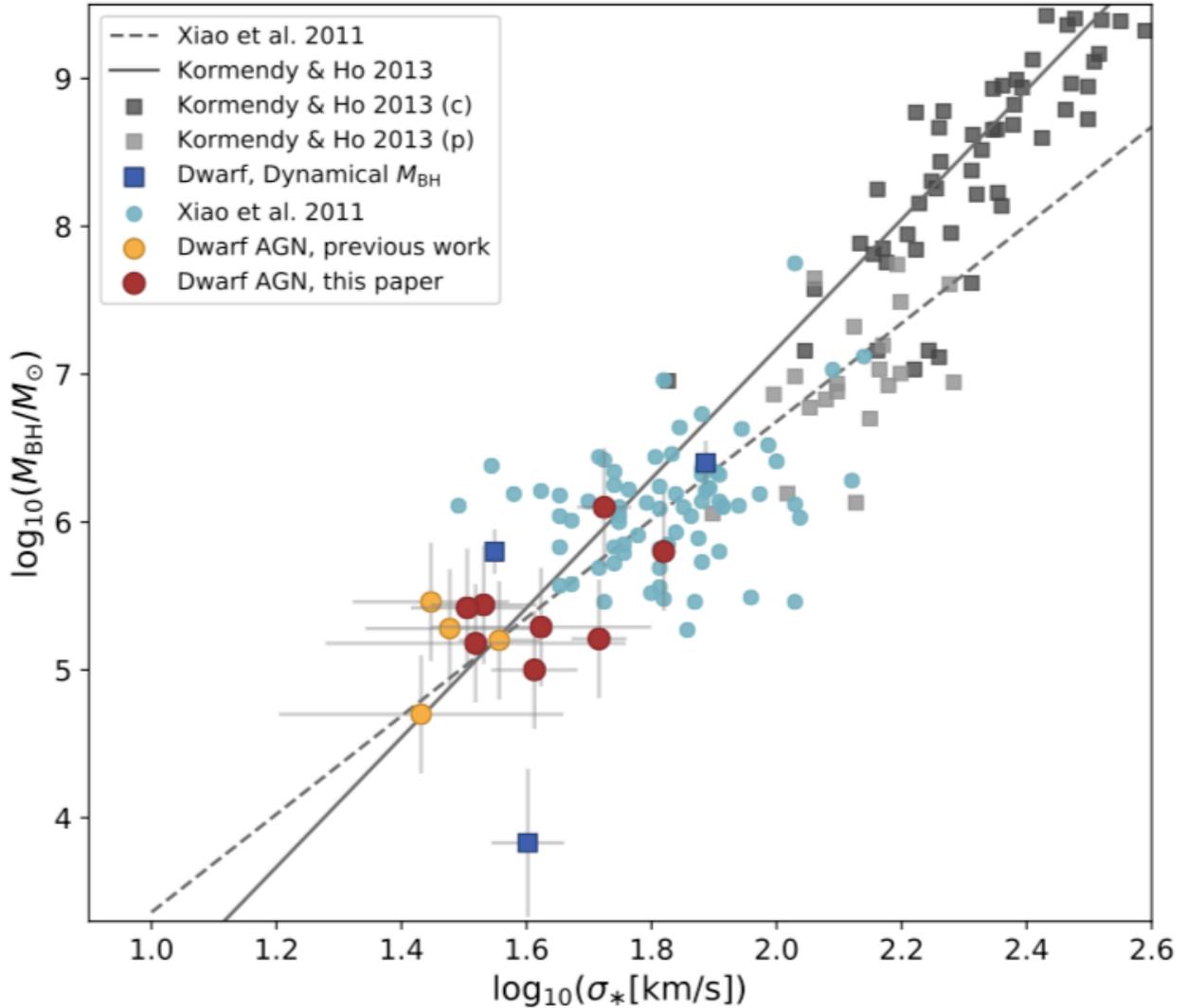

Fig. 5. Black hole mass ($M_{BH}$) versus stellar velocity dispersion ($\sigma_*$). Gray squares are from the compilation of Kormendy & Ho (2013); dark gray squares show galaxies with classical bulges and light gray squares show galaxies with pseudo-bulges. Light blue circles show data for low-mass AGN from Xiao et al. (2011). Active dwarf galaxies with stellar velocity dispersion measurements measured in Baldassare et al. (2020) are shown as red circles; active dwarf galaxies with previously existing data are shown in orange. Dwarf galaxies with dynamical black hole mass estimates are shown as dark blue squares. Also shown are the fits to the $M_{BH}$ - $\sigma_*$ from Xiao et al. (2011; dashed line) and Kormendy & Ho (2013; solid line). Credit: Figure and caption adapted from Baldassare et al. 2020, ApJ Letters, 898, L3, reproduced with permission.

**Gravitational waves.**

Gravitational waves are ripples in the space-time that propagate at the speed of light. They are generated by massive accelerated objects such as two merging black holes. In a black hole binary, gravitational waves are emitted during the inspiral phase, the merger, and the final ring-down. This was the case of the first detection of gravitational waves by the LIGO and Virgo interferometers in 2015, when two black holes of 29 and 36 $M_\odot$ merged into a new black hole of 62 $M_\odot$ (Abbott et al. 2016). Since then, tens of gravitational wave observations resulting from the merger of two stellar-mass black holes have been reported (Fig. 1). One of such black hole binaries resulted in a final black hole of 142 $M_\odot$, thus in the IMBH regime (Abbott et al. 2020). The detection of gravitational waves from SMBH binaries will have to await the next generation of space-based interferometers (e.g. Amaro-Seoane et al. 2018).

## Key Research Findings

**Universality of black hole accretion.**

The process of black hole accretion and its connection with the ejection of large-scale outflows (i.e. jets) is predicted to be a universal mechanism covering the whole black hole mass scale. This implies that the same disc-jet connection governing stellar-mass black holes does also take place in SMBHs (e.g. Falcke et al. 2004). Observational evidence for this universality of black hole accretion comes from the finding of a correlation between X-ray luminosity (proxy of accretion flow), radio luminosity (proxy of jet emission), and black hole mass extending from stellar-mass to SMBHs (the so-called Fundamental Plane of Black Hole Accretion; see Fig. 4; e.g. Merloni et al. 2003; Plotkin et al. 2012). The Fundamental Plane of Black Hole Accretion can be used to estimate black hole masses when nuclear radio and X-ray luminosities are available (e.g. Mezcua et al. 2015, 2018c: Gultekin et al. 2019).

**Black hole-galaxy co-evolution.**

Another key finding in black hole studies is that of a correlation between the mass of SMBHs and some host galaxy properties, such as bulge luminosity, stellar mass, or stellar velocity dispersion (see review by e.g. Kormendy & Ho 2013), which implies that SMBHs and their host galaxies co-evolve (see Fig. 5). While this synchronized black hole-galaxy growth was originally found for SMBHs of $> 10^6$ M$_\odot$, recent studies have extended these correlations to the IMBH regime, where a flattening or large scatter seems to be observed (e.g. Martin-Navarro & Mezcua 2018; Baldassare et al. 2020; Greene et al. 2020). Feedback from the AGN onto the host galaxy is thought to be the main mechanism behind such black hole-galaxy evolution, via winds or jets that impact the formation of stars in the host galaxy and thus regulate its growth. Observational evidence for AGN feedback in massive galaxies comes from the finding that the radio jets of SMBHs fill the large-scale X-ray cavities of some galaxy clusters (e.g. McNamara et al. 2000). The effects of AGN feedback in dwarf galaxies is not so clear, but recent simulations (e.g. Sharma et al. 2020 and references therein) and observations (e.g. Mezcua et al. 2019; Liu et al. 2020) indicate it could be as significant as in massive galaxies.

## Future Directions

The next generation of observational facilities will provide a major step forward in our understanding of black hole formation and evolution and the physics behind black hole accretion. Future radio observatories such as the Square Kilometer Array (SKA) and X-ray satellites such as Athena or the mission-concept Lynx will provide a complete coverage of the phases that stellar-mass black hole transients undergo (i.e. the disk-jet coupling mechanism) and be able to detect SMBHs and their seed IMBHs out to z=10 (e.g. Yu et al. 2015; Mezcua et al. 2018a; Whalen et al. 2020). In the new era of gravitational-wave astronomy, the Einstein Telescope will capture the gravitational-wave signal from millions of coalescing stellar binary black holes detectable out to z~15 and up to masses of $10^3$ M$_\odot$ characteristic of (growing) black hole seeds, while LISA will detect the gravitational-wave signal from massive binary black hole coalescences (~ $10^4$ M$_\odot$ to ~$10^7$ M$_\odot$ ) across all cosmic ages (e.g. Amaro-Seoane et al. 2018). A future of exciting discoveries awaits us.

## See Also

→ Active Galactic Nuclei

→ Black Holes

→ Galaxy

→ Globular Cluster

→ Radio Astronomy

→ [X-ray binaries](#)

# References

Abbott B. P., et al., 2016, Physical Review Letters, 116, 241103
[ADS](#)

Abbott B. P., et al., 2020, Physical Review Letters, 125, 101102
[ADS](#)

Amaro-Seoane P., 2018, Physics Review D, 98, 063018
[ADS](#)

Baldassare V. F., et al. 2016, ApJ, 829, 57
[ADS](#)

Baldassare V. F., et al., 2020, ApJ Letters, 898, L3
[ADS](#)

Barrows R. S., et al. 2019, ApJ, 882, 181
[ADS](#)

Carr, B.; Kühnel, F., 2020, Annual Review of Nuclear and Particle Science, 70, 355
[ADS](#)

Casares J., Jonker P. G., 2014, Space Science Reviews, 183, 223
[ADS](#)

Chilingarian I. V., et al. 2018, ApJ, 863, 1
[ADS](#)

Davis T. A., et al., 2020, MNRAS, 496, 4061
[ADS](#)

Falcke H., at al. , 2004, A&A, 414, 895
[ADS](#)

Gallo E., et al., 2010, ApJ, 714, 25
[ADS](#)

García-Bellido, J., 2019, Philosophical Transactions of the Royal Society A: Mathematical, Physical and Engineering Sciences, 377, 2161
ADS

Gebhardt K., et al., 2005, ApJ, 634, 1093
ADS

Ghez A. M., et al., 2008, ApJ, 689, 1044
ADS

Gillessen S., et al. 2009, ApJ, 692, 1075
ADS

Greene J. E., et al., 2020, ARA&A, 58, 257
ADS

Gültekin K., et al., 2019, ApJ, 871, 80
ADS

Kormendy J., Ho L. C., 2013, ARA&A, 51, 511
ADS

Liu W., et al., 2020, ApJ, 905, 166
ADS

Lützgendorf N., et al. 2013, A&A, 552, A49
ADS

Martín-Navarro I., Mezcua M., 2018, ApJ Letters, 855, L20
ADS

McNamara B. R., et al., 2000, ApJ Letters, 534, L135
ADS

Merloni A.; Heinz S.; di Matteo T., 2003, MNRAS, 345, 1057
ADS

Mezcua M. 2017, International Journal of Modern Physics D, 26, 1730021
ADS

Mezcua M., Domínguez Sánchez H., 2020, ApJ Letters, 898, L30
[ADS](#)

Mezcua M., et al., 2015, MNRAS, 448, 1893
[ADS](#)

Mezcua M., et al., 2016, ApJ, 817, 20
[ADS](#)

Mezcua M., et al., 2018a, MNRAS, 474, 1342
[ADS](#)

Mezcua M., et al., 2018b, MNRAS, 478, 2576
[ADS](#)

Mezcua M., et al., 2018c, MNRAS, 480, L74
[ADS](#)

Mezcua M., et al., 2019, MNRAS, 488, 685
[ADS](#)

Pasham D. R., et al., 2014, Nature, 513, 74
[ADS](#)

Plotkin R. M., et al., 2012, MNRAS, 419, 267
[ADS](#)

Rees 1978, The Observatory, 98, 210
[ADS](#)

Reines A. E., et al., 2013, ApJ, 775, 116
[ADS](#)

Remillard R. A., McClintock J. E., 2006, ARA&A, 44, 49
[ADS](#)

Rubin, S. G., Sakharov, A. S., Khlopov, M. Y. 2001, Journal of Experimental and Theoretical Physics, 92, 921
[ADS](#)

Sharma R.S., et al., 2020, ApJ, 897, 103
[ADS](#)

Stern D., et al., 2012, ApJ, 753, 30
ADS

The EHT Collaboration, et al., 2019, ApJ Letter, 875, L1
ADS

Volonteri, M.; Habouzit, M.; Colpi, M., 2021, Nature Review Physics, 3, 732
ADS

Wang F., et al., 2021, ApJ Letters, 907, L1
ADS

Whalen D. J., et al., 2020, ApJ Letters, 896, 45
ADS

Wrobel J. M., Nyland K. E., 2020, ApJ, 900, 134
ADS

Xiao T., et al., 2011, ApJ, 739, 28
ADS

Yu W. et al., 2015, "Advancing Astrophysics with the Square Kilometre Array", Proceedings of Science, 66
ADS